\documentclass[runningheads]{llncs}
\usepackage[T1]{fontenc}
\usepackage{graphicx}
\usepackage{caption}
\usepackage{subcaption}
\usepackage{amsmath}
\usepackage{amssymb}
\usepackage{textcomp}
\usepackage{booktabs}
\usepackage[justification=justified]{caption}
\makeatletter  
\newif\if@restonecol  
\makeatother

\makeatletter  
\newif\if@restonecol  
\makeatother

\usepackage[linesnumbered,ruled,vlined]{algorithm2e}
\usepackage{algpseudocode}  
\usepackage{color} 
\usepackage{float}
\usepackage{hyperref} 
\usepackage{url}
\usepackage{multirow}
\usepackage{orcidlink}
\usepackage{bbding}
\usepackage{booktabs}
\usepackage{appendix}

\begin{document}
	\begin{sloppypar}
		
		\title{ChainsFormer: A Chain Latency-aware Resource Provisioning Approach for Microservices Cluster}
		%
		%
		\author{Chenghao Song\inst{1}\orcidlink{0000-0002-4570-2722} \and
		Minxian Xu\inst{1}(\Envelope)\orcidlink{0000-0002-0046-5153} \and
		Kejiang Ye\inst{1}\orcidlink{0000-0001-8985-2792} \and
		Huaming Wu \inst{2}\orcidlink{0000-0002-4761-9973} \and
		Sukhpal Singh Gill \inst{3}  \orcidlink{0000-0002-3913-0369} \and   
		Rajkumar Buyya\inst{4}\orcidlink{0000-0001-9754-6496} \and
		Chengzhong Xu\inst{5}\orcidlink{0000-0001-9480-0356} 
		}

		\authorrunning{C. Song et al.}
		\titlerunning{Latency-aware Resource Provisioning for Microservices Cluster}
		%
		\institute{Shenzhen Institute of Advanced Technology, Chinese Academy of Sciences, China \\
		\email{\{ch.song, mx.xu, kj.ye\}@siat.ac.cn}
		\and
		Tianjin University, Tianjin, China
		\email{whming@tju.edu.cn}
		\and
		 Queen Mary University of London, London, UK
		 \email{s.s.gill@qmul.ac.uk}\\
		 \and
		Cloud Computing and Distributed Systems (CLOUDS) Lab, School of
		Computing and Information Systems, The University of Melbourne, Australia   
		\email{rbuyya@unimelb.edu.au} \\
		 \and
		State Key Lab of IoTSC, University of Macau, Macau, China
		\email{czxu@um.edu.mo}
		}

		\let\oldmaketitle\maketitle
		\renewcommand{\maketitle}{\oldmaketitle\setcounter{footnote}{0}}

		\maketitle              

		\begin{abstract}
			The trend towards transitioning from monolithic applications to microservices has been widely embraced in modern distributed systems and applications. This shift has resulted in the creation of lightweight, fine-grained, and self-contained microservices. Multiple microservices can be linked together via calls and inter-dependencies to form complex functions. One of the challenges in managing microservices is provisioning the optimal amount of resources for microservices in the chain to ensure application performance while improving resource usage efficiency. This paper presents \textit{ChainsFormer}, a framework that analyzes microservice inter-dependencies to identify critical chains and nodes, and provision resources based on reinforcement learning. To analyze chains, ChainsFormer utilizes light-weight machine learning techniques to address the dynamic nature of microservice chains and workloads. For resource provisioning, a reinforcement learning approach is used that combines vertical and horizontal scaling to determine the amount of allocated resources and the number of replicates. 
			\color{black}We evaluate the effectiveness of \textit{ChainsFormer} using realistic applications and traces on a real testbed based on Kubernetes. Our experimental results demonstrate that \textit{ChainsFormer} can reduce response time by up to 26\% and improve processed requests per second by 8\% compared with state-of-the-art techniques.\color{black} 
			
		\end{abstract}

		\vspace{-0.3cm}

		\section{Introduction}

Microservice architecture is a popular approach for designing and developing modern applications. It involves breaking down monolithic applications into smaller, fine-grained components that can work together to provide services for users \cite{XuContainerSurvey}. This approach allows development teams to focus on implementing different microservices, thereby speeding up the development process. Additionally, microservices can be updated or upgraded independently, making maintenance efforts more manageable. To ensure reliability and performance, microservices can be scaled and operated individually, depending on workload fluctuations and environmental variance. 

Despite their independence, microservices are not entirely self-contained. Communication-based dependencies, such as remote procedure calls, exist between different microservices \cite{luo2021characterizing}. These dependencies can represent how requests are processed among different microservices. Based on these dependencies, microservices can be combined into chains to fulfill complex services. The length of a chain can vary from several nodes to tens of nodes. A single microservice, such as a database-related service, can also be shared by multiple chains to support the formation of different services. Additionally, microservice chains can be dynamic, scaling in or out as needed to accommodate new microservices. Given these features of microservices and the resource usage fluctuations, it is challenging to precisely pre-configure the amount of resources, provision and scale resources when deploying microservices in clusters. 

Traditional approaches for improving application performance often rely on over-provisioning and autoscaling, which involve allocating more CPU and memory resources to microservices. These approaches typically use performance models, simple heuristics, static thresholds, or machine learning algorithms. However, these approaches have several limitations.
Firstly, accurate performance models and efficient heuristic-based scheduling policies require significant manual efforts and training, which are infeasible for large-scale microservices with a large number of configurable parameters.
Secondly, machine learning (ML) based approaches, such as support vector machines, rely on centralized graph databases, which can lead to scalability issues and inefficient scheduling when microservice chains are updated.
Therefore, alternative approaches are needed to address these limitations and enable effective management of microservices. 

This paper presents a solution to the limitations of traditional approaches with \textit{ChainsFormer}, a \underline{chain} latency-aware re\underline{s}ource provisioning framework \underline{for} \underline{m}icros\underline{er}vices cluster based on chain feature analysis. \textit{ChainsFormer} dynamically scales CPU and memory resources to microservices to ensure high-quality service. This framework utilizes online telemetry data, including requests information, application running data, and hardware resource usage, to capture the system state. By leveraging ML and reinforcement learning (RL) models, \textit{ChainsFormer} can adapt to variances in the system and reduce the need for manual efforts. Overall, \textit{ChainsFormer} provides an effective solution for managing microservices with a high degree of automation and accuracy. 

\color{black}To efficiently manage the dynamic nature of microservice chains and adapt to changes quickly, \textit{ChainsFormer} employs various techniques. It first identifies the critical chain using the calling graph and utilizes a decision tree to find the critical node that has a significant influence on microservice performance. This approach avoids the limitations of heavy machine learning techniques and centralized graph databases, which struggle with dynamic changes.
Additionally, \textit{ChainsFormer} utilizes RL to make efficient and optimized decisions regarding vertical and horizontal scaling. These decisions include when to conduct scaling actions, which microservice should be scaled with resources, and how many resources of each type should be scaled. Furthermore, these decisions can be further optimized through RL with updated decisions, resulting in even more efficient resource provisioning.\color{black}

To evaluate the effectiveness of \textit{ChainsFormer}, we deployed representative microservice applications on Kubernetes, which is the state-of-the-art container orchestration platform. We compared \textit{ChainsFormer} with three state-of-the-art baselines and used realistic traces from Alibaba to measure application performance and response time.
Our results show that \textit{ChainsFormer} outperforms the baselines in terms of application performance and response time. 
These findings demonstrate the effectiveness of \textit{ChainsFormer} in providing efficient resource provisioning and management for microservice chains.

In summary, we make the following key \textbf{contributions}: 
\begin{itemize}
	\color{black}
	\item  We present the design of a framework that aims to handle the dynamic changes in microservice applications by identifying critical chains and nodes.  
	\item  We propose an RL-based approach for combining vertical and horizontal scaling to make decisions on efficient resource provisioning, which uses historical data for offline training and makes online decisions based on system states. 
	\item We develop the designed framework on top of Kubernetes platform. Using realistic workload traces and real-world microservice, we demonstrate the efficiency of \textit{ChainsFormer} compared to the state-of-the-art baselines. 
\end{itemize}

\section{Related Work}

\label{sec:Background}

In this section, we will discuss the current state-of-the-art techniques that are designed to address the challenges of resource provisioning and autoscaling in microservices, in order to meet the desired quality of service levels.

\begin{table*}[t]
	\centering
	\caption{Comparison of related work}
	\label{tab:related_work}
	\resizebox{0.95\textwidth}{!}{%
		\begin{tabular}{@{}ccccccc@{}}
			\hline
			\textbf{Approach} &
			\textbf{Autoscaling} &
			\textbf{\begin{tabular}[c]{@{}c@{}}Workloads \\ Prediction \end{tabular}}&
			\textbf{\begin{tabular}[c]{@{}c@{}}Machine Learning \\ based Resource \\ Provisioning\end{tabular}} &
			\textbf{ \begin{tabular}[c]{@{}c@{}}Chain \\ Analysis\end{tabular}} &
			\textbf{\begin{tabular}[c]{@{}c@{}}Quick Adaption \\ to Dynamic Chains\end{tabular}} &
			\textbf{ SLO-awareness} \\ \hline
			\textbf{Sage} \cite{Sage2021}                 &              & $\checkmark$ & $\checkmark$ & Partial      &              & $\checkmark$ \\ \hline
			\textbf{Firm} \cite{qiu2020firm}              & $\checkmark$ &              & $\checkmark$ & $\checkmark$ &              & $\checkmark$ \\ \hline
			\textbf{Parslo} \cite{Parslo2021}             & $\checkmark$ &              &              & $\checkmark$ &              & $\checkmark$ \\ \hline
			\textbf{PEMA}  \cite{PEMA2022}                & $\checkmark$ &              &              &              &              & $\checkmark$ \\ \hline
			\textbf{Autopilot} \cite{rzadca2020autopilot} & $\checkmark$ & $\checkmark$ & $\checkmark$ &              &              & $\checkmark$ \\ \hline
			\textbf{Sinan}   \cite{Sinan2021}             & $\checkmark$ &              & $\checkmark$ & $\checkmark$ &              & $\checkmark$ \\ \hline
			\textbf{Seer}  \cite{gan2019seer}             &              &              & $\checkmark$ & $\checkmark$ & $\checkmark$ & $\checkmark$ \\ \hline
			\textbf{CoScal}  \cite{XuCoScal}             &    $\checkmark$          &    $\checkmark$          & $\checkmark$ & &  & $\checkmark$ \\ \hline
			\textbf{ChainsFormer (ours)}                   & $\checkmark$ & $\checkmark$ & $\checkmark$ & $\checkmark$ & $\checkmark$ & $\checkmark$ \\ \hline
		\end{tabular}%
	}
\end{table*}

\textbf{Resource Provisioning for Microservices.} Sage \cite{Sage2021} aims to perform root cause analysis in microservice-based systems by utilizing causal Bayesian networks to identify the underlying reason for service level objective (SLO) violations. After identifying the root cause, Sage initiates autoscaling actions to mitigate the issue. One of the advantages of Sage is that it only requires lightweight tracking and is suitable for large-scale deployments. However, a major limitation of Sage is its heavy reliance on pre-trained machine learning models.
Seer \cite{gan2019seer} employs deep learning models to predict quality of service (QoS) violations and dynamically adjusts allocated resources to each microservice to prevent such violations. It is particularly suitable for scenarios with frequent service updates and requires a large amount of tracking data. However, the accuracy of detection can be affected by significant application changes.
Parslo \cite{Parslo2021} is a gradient descent-based approach that assigns partial SLOs to nodes in a microservice to provide resource configuration solutions quickly. One of Parslo's main advantages is its ability to achieve a globally optimal solution for large-scale services that have already been deployed. However, Parslo is limited in its support for only certain types of Directed Acyclic Graphs, and its performance may not be guaranteed in all circumstances.
PEMA \cite{PEMA2022} uses iterative feedback-based tuning to optimize resource allocation to meet SLO requirements. Compared to other approaches, PEMA is lightweight and does not require any offline experiments or pre-training. However, PEMA's performance may be poor during resource update intervals, and its inability to capture the dependencies between microservices due to the lack of pre-training may limit its effectiveness.
The fundamental limitation of this line of work is that they do not consider features of microservice chain, which can lead to inefficient actions and performance degradation. 

\textbf{Micoservice Autoscaling.} Autopilot \cite{rzadca2020autopilot} utilizes ML algorithms to analyze historical data on prior executions and performs a set of finely-tuned heuristics to adjust a job's resource requirements while it is running. The benefit of Autopilot is its ability to modify resource requirements on-the-fly, allowing it to adapt to changing workload demands. However, Autopilot's conservative approach can lead to overprovisioning and resource wastage.
Sinan \cite{gan2019seer} leverages a set of machine learning models to determine the performance impacts of microservice dependencies and allocate appropriate resources for each tier. Sinan is an explainable approach and can be used for complex microservices, while it only monitors CPU resources and does not provide auto-tuning capabilities.
CoScal \cite{XuCoScal} leverages data-driven decisions and enables multi-faceted scaling based on reinforcement learning. CoScal utilizes gradient recurrent units to accurately predict workloads, which assists in achieving efficient scaling. However, one limitation is that the model re-training required for adapting to new applications can be costly.
FIRM \cite{qiu2020firm} is a system that utilizes online telemetry data and machine learning methods to adaptively detect and locate microservices that lead to SLO violations. It can make decisions based on reinforcement learning to mitigate SLO violations via fine-grained and dynamic resource provisioning. FIRM proposes a two-level ML framework to locate critical microservice paths and nodes. However, FIRM has certain limitations. The scalability of centralized graph databases is limited, and it cannot handle transient SLO violations that occur within an interval shorter than the minimum interval due to the heavy ML techniques.

\color{black}The comparisons between \textit{ChainsFormer} and other relevant work are presented in Table \ref{tab:related_work}. Our work is most similar to CoScal and FIRM. However, there are notable differences between them. Firstly, CoScal is deployed on Docker Swarm, while \textit{ChainsFormer} is designed specifically for Kubernetes. Secondly, CoScal does not incorporate chain analysis for resource management, whereas \textit{ChainsFormer} leverages chain analysis techniques to optimize resource allocation within microservice chains. Thirdly, both CoScal and \textit{ChainsFormer} employ reinforcement learning, but \textit{ChainsFormer} utilizes the SARSA algorithm, which allows for faster convergence by updating Q-values based on the current policy.
In comparison to FIRM, \textit{ChainsFormer} employs lightweight ML techniques to handle transient SLO violations in microservice chains, a task that FIRM does not address due to its heavy ML models and the associated high costs of model re-training. Additionally, \textit{ChainsFormer} does not require a centralized graph database like FIRM, which enhances its scalability by avoiding a central bottleneck caused by large amounts of data. In \textit{ChainsFormer}, runtime data is stored on worker nodes and only fetched by the central node when model training or retraining is required, significantly reducing the overhead on the central node. \color{black}


\color{black}

\section{The ChainsFormer Framework}

\label{sec:Policy}

\begin{figure*}[t]
	\vspace{-0.1cm}\centering
	\begin{subfigure}[b]{0.45\linewidth}
		\centering
		\includegraphics[width=1\linewidth]{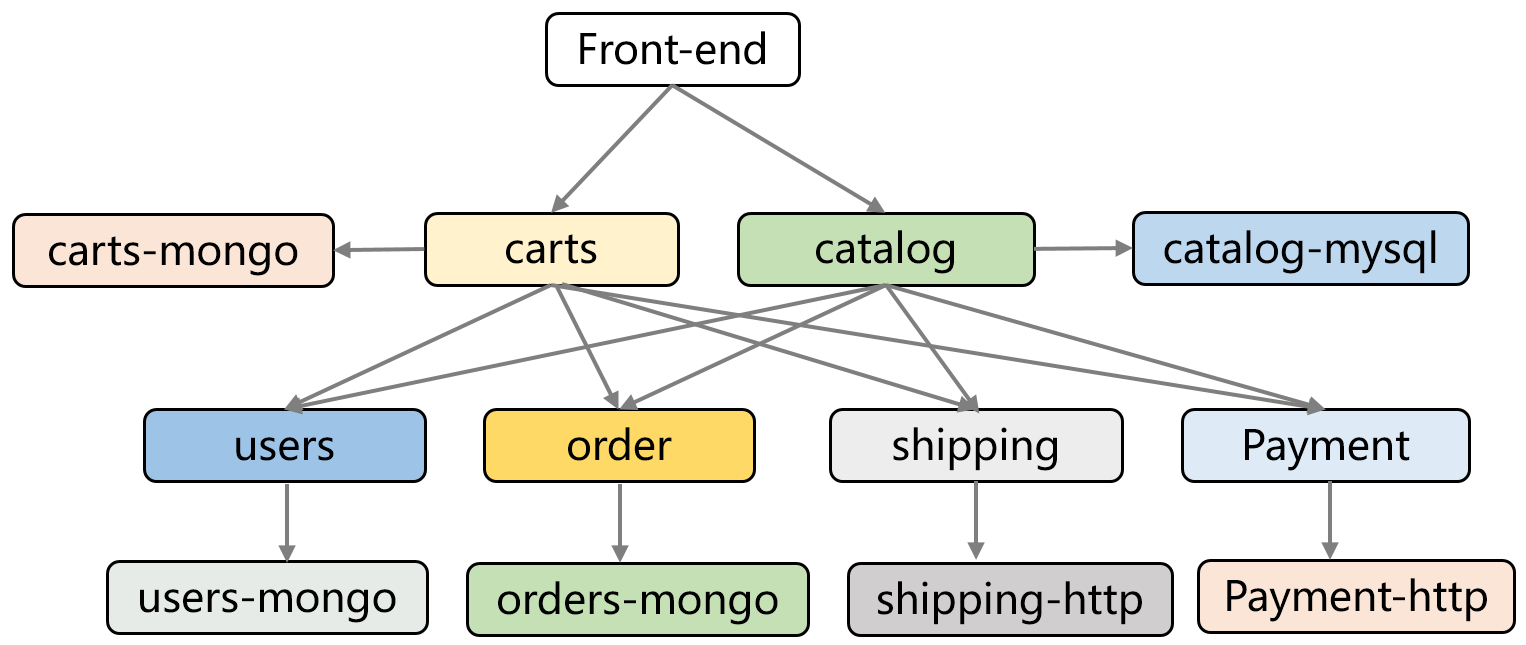}
		\caption{} 
		\label{fig:topo}
	\end{subfigure}
	\begin{subfigure}[b]{0.45\linewidth}
		\centering
		\includegraphics[width=1\linewidth]{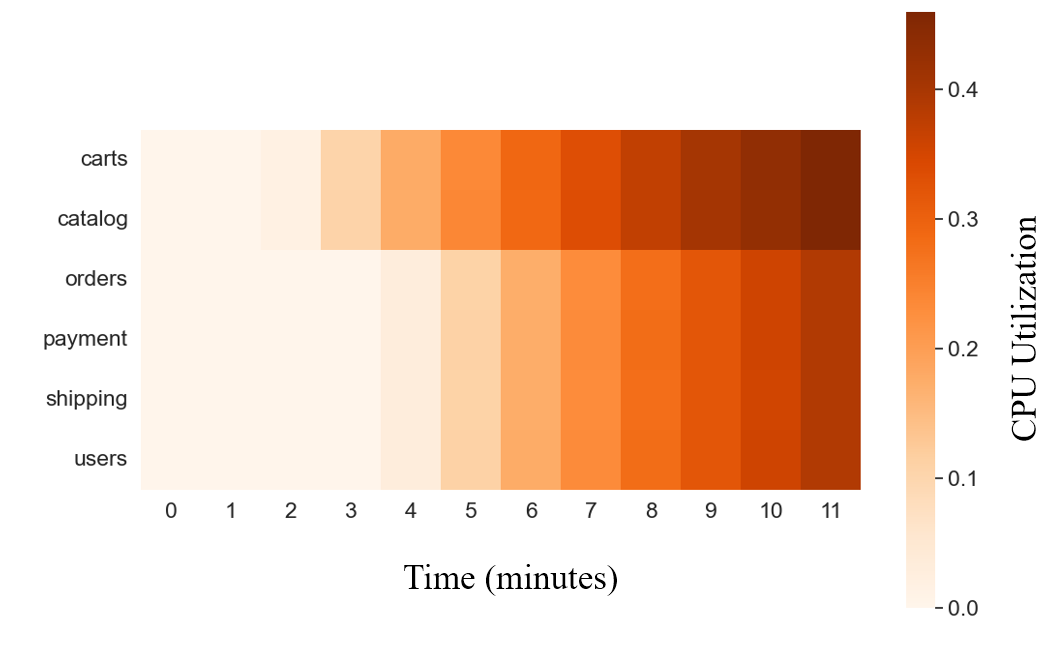}
		\caption{} 
		\label{fig:heatmap}
	\end{subfigure}
	\vspace{-0.1cm}
	\caption{(a) Microservice graph structure of Sock Shop application. (b) CPU utilization of each microservice from top tier to bottom tier. When workloads increase, the CPU utilization of all microservices increases.}
	\label{fig:motivation}
\end{figure*}

\color{black}To motivate our design, we deployed the Sock Shop application\footnote{Sock Shop: A Microservices Demo Application. https://microservices-demo.github.io/} to observe how different microservices react to changes in workloads by monitoring utilization usage. \color{black}As shown in Fig. \ref{fig:topo}, a request sent to the Sock Shop application can be distributed to different microservices from front-end to back-end tiers. The processing of a request can form different calling chains, for example, a request can go through different chains to complete different functionalities, e.g. checking items under a user account (front-end $\rightarrow$ carts $\rightarrow$ users), or paying for an item (front-end $\rightarrow$ catalog $\rightarrow$ payment). As shown in Fig. \ref{fig:heatmap}, workloads increase from 0-110 requests per second during 0-11 minutes (requests per second is increased with 10 after each minute), and the CPU utilization also increases for all microservices, while the resource usage propagation among the nodes in a chain is not consistent. Thus, to achieve efficient resource provisioning of microservices, the scheduler should consider the features of the microservice chain properly.

\begin{figure}[!ht]
	\centering
	\includegraphics[width=0.99\linewidth]{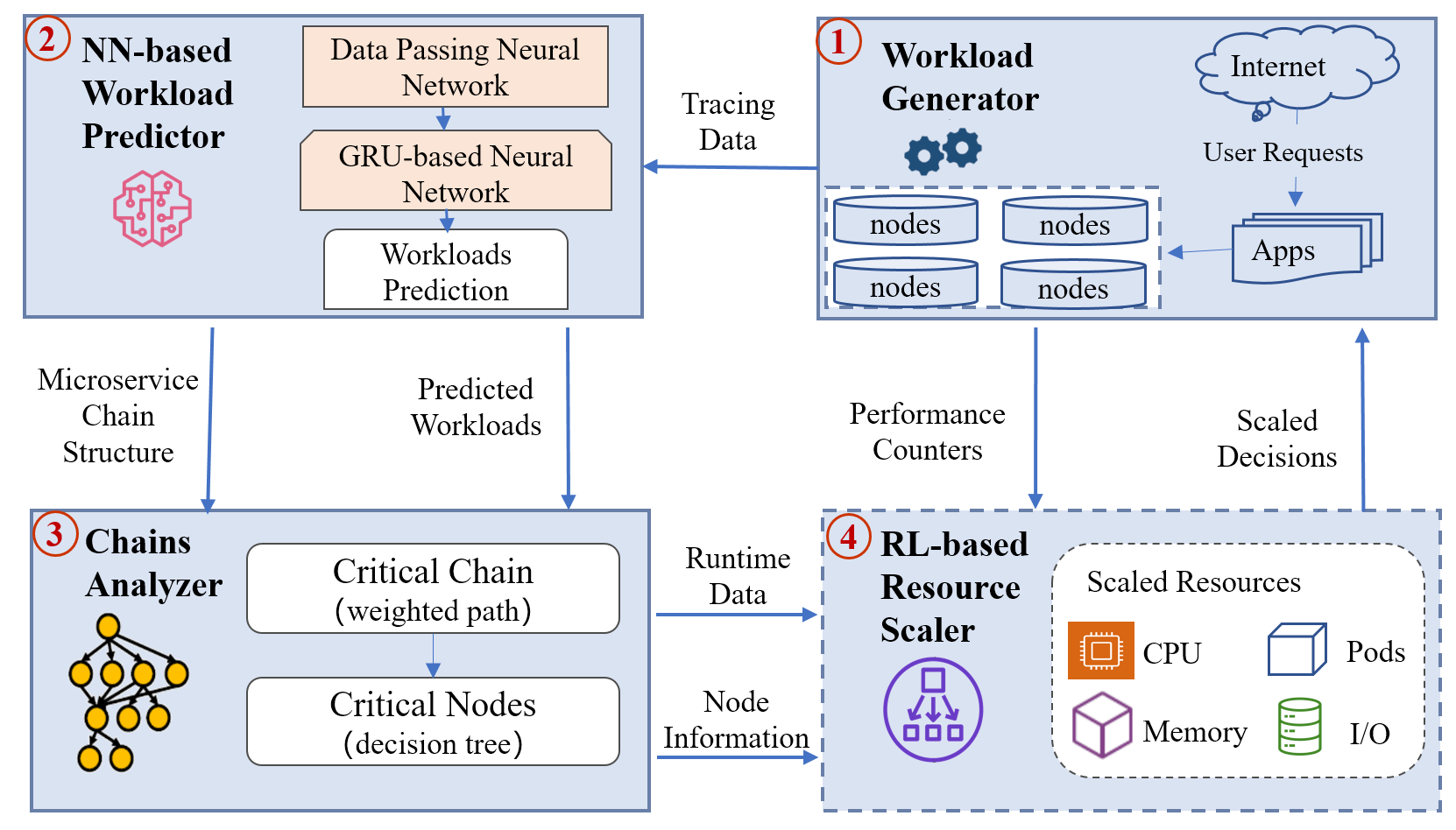}
	\vspace{-0.1cm}
	\caption{Framework of \textit{ChainsFormer}}
	\label{fig:MAPEK}
\end{figure}

To address the above observations, we propose the overview architecture design of \textit{ChainsFormer} as shown in Fig. \ref{fig:MAPEK} and the key designs are as below:
\begin{itemize}
	\item  \textit{ChainsFormer} first processes the incoming requests from users via Workload Generator by recording the number of requests and extracting the tracing data and performance counters. 
	
	\item 	To make the resource provisioning more efficient, \textit{ChainsFormer} applies the neural network-based prediction algorithm to estimate future workloads. 
	
	\item 	\textit{ChainsFormer} detects SLO violations and utilizes real-time data to dynamically identify critical chains and locate critical nodes that result in SLO violations. To support the quick adaption to the dynamic changes in chains, \textit{ChainsFormer} includes an auto-adaptor that can quickly detect the changes. 
	
	\item 	\textit{ChainsFormer} analyzes the telemetry data collected by Workload Generator and node information identified by Chains Analyzer, and makes scaling decisions to provision resources for critical nodes. The decision is made automatically on the Kubernetes cluster by an RL-based resource scaler, which considers resource utilization, performance metrics, and future workloads.
	
\end{itemize}

\subsection{Workload Generator}
\label{sec:workloadsgen}

Workload Generator module in \textit{ChainsFormer} is responsible for processing the raw workload trace to make fit with other modules, e.g. extracting the key information of workloads (e.g. timestamp and user id) and providing initial analyses for the workloads. Based on the required functionalities, workloads are distributed to different microservices that are deployed on different work nodes in the microservices cluster. For example, we have observed that the workloads of the Sock Shop application are distributed to Front-end (45.5\%), Order (22.7\%), Carts (22.7\%), Catalog (5.7\%) and Random item (3.4\%) with different percentages. 
The processed workloads are also regularly stored in log files for workloads prediction, and the performance counters that indicate system performance are provided to resource scalers for autoscaling microservices.  

\begin{figure*}[t]
	\centering
	\begin{subfigure}[b]{0.24\linewidth}
		\centering
		\includegraphics[width=1\linewidth]{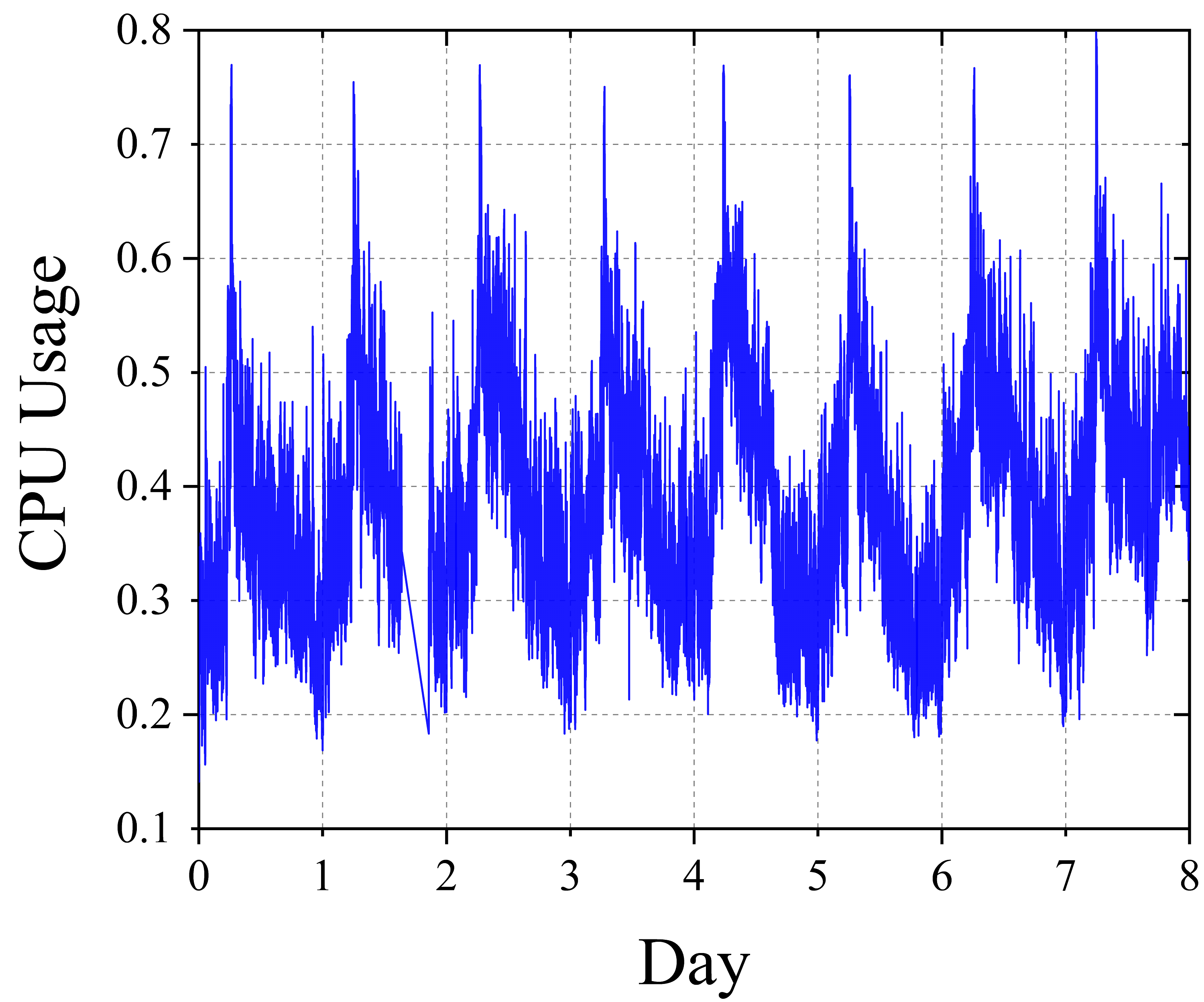}
		\caption{} 
		\label{fig:Alibaba_per-day_workload_old}
	\end{subfigure}
	\begin{subfigure}[b]{0.24\linewidth}
		\centering
		\includegraphics[width=1\linewidth]{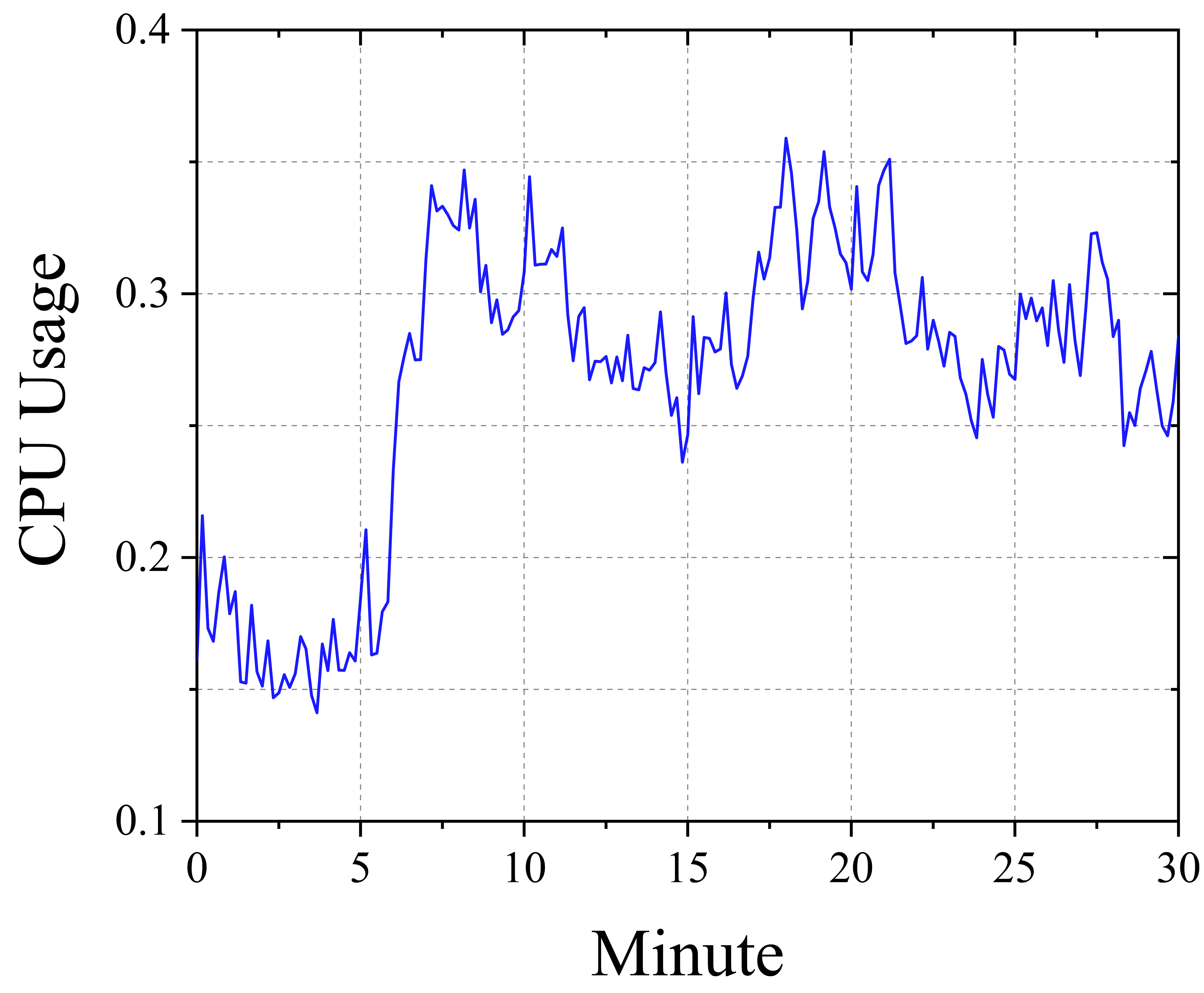}
		\caption{} 
		\label{fig:Alibaba_per-minute_workload_old}
	\end{subfigure}
	\begin{subfigure}[b]{0.24\linewidth}
		\centering
		\includegraphics[width=1\linewidth]{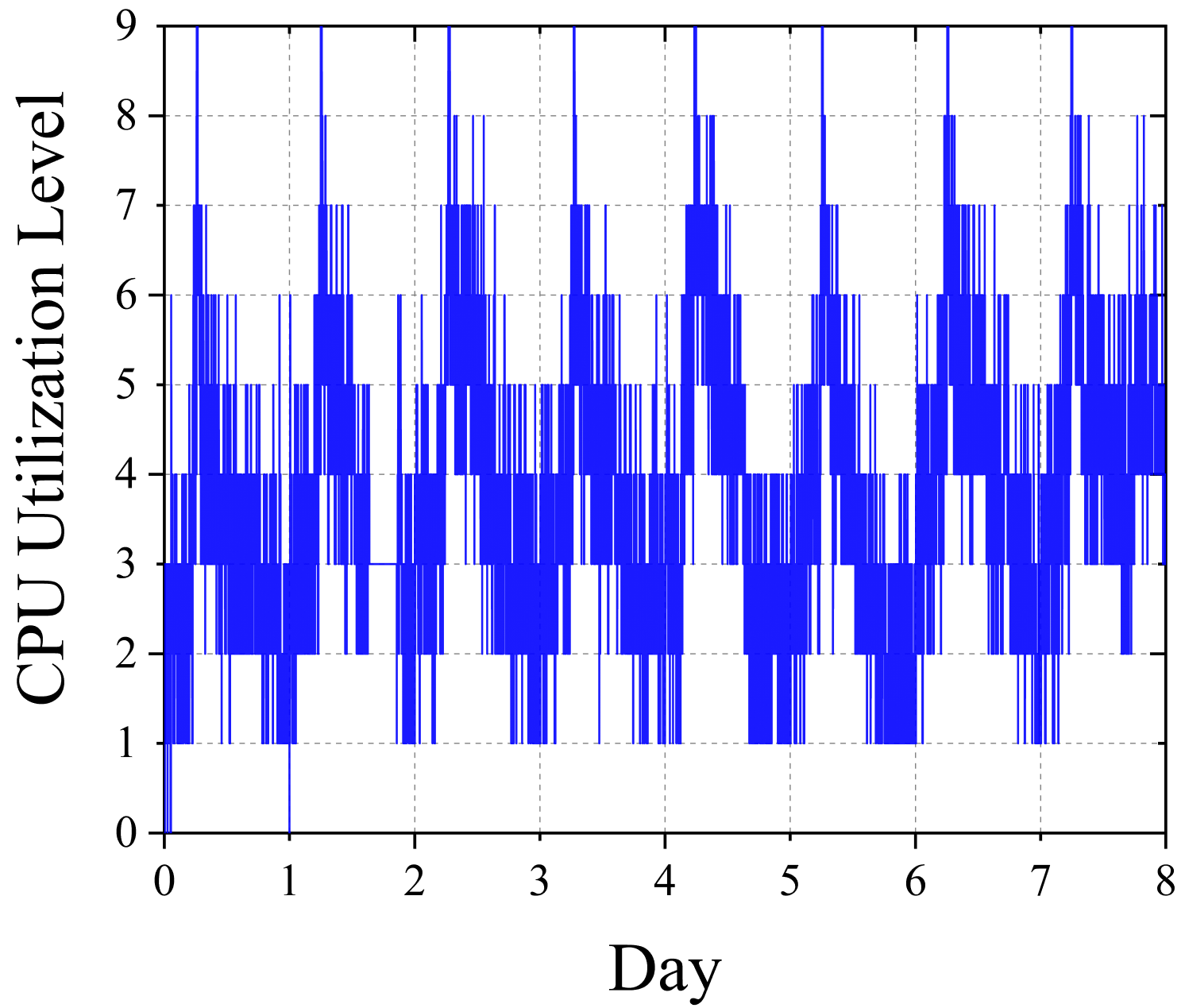}
		\caption{} 
		\label{fig:Alibaba_per-day_workload}
	\end{subfigure}
	\begin{subfigure}[b]{0.24\linewidth}
		\centering
		\includegraphics[width=1\linewidth]{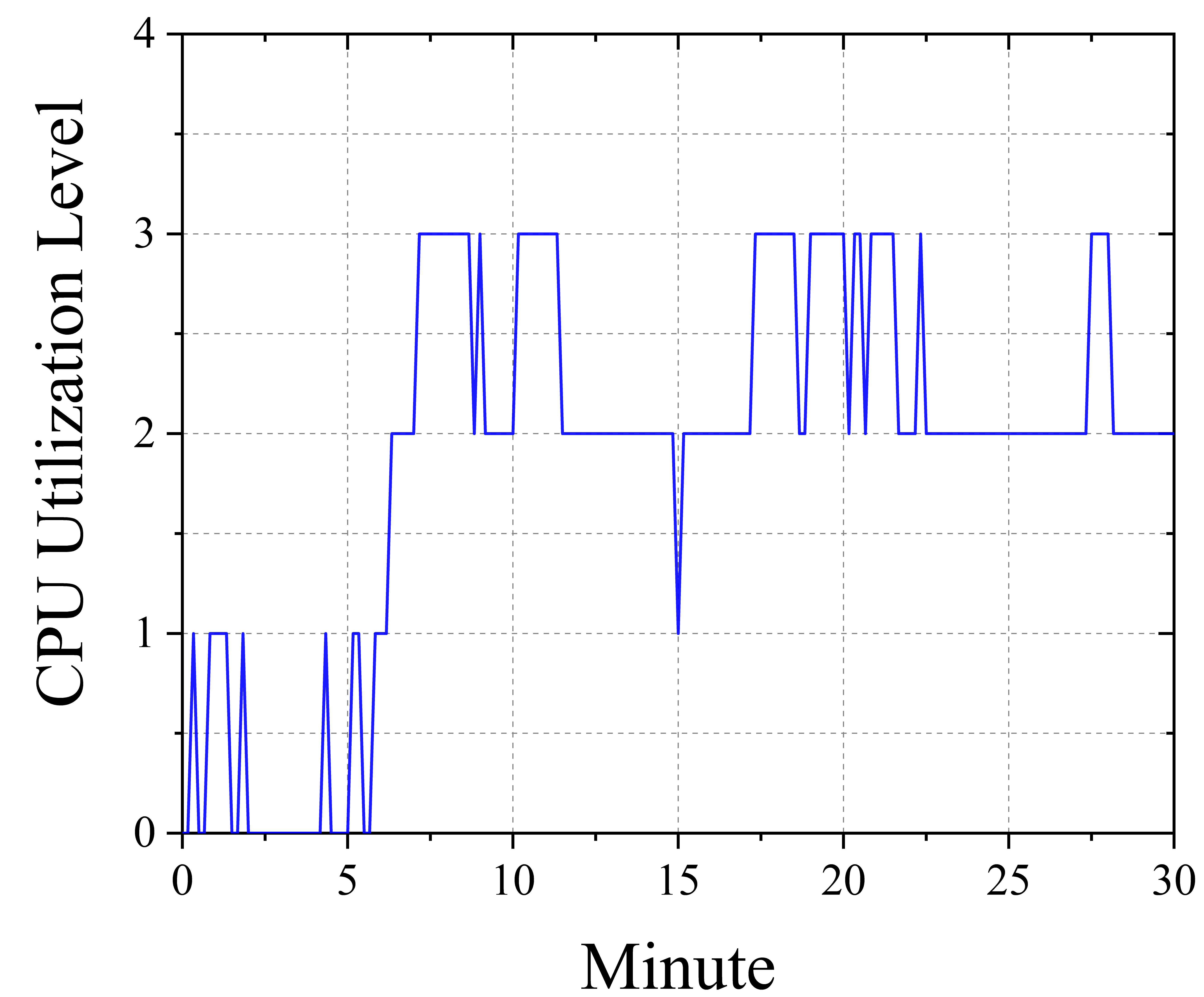}
		\caption{} 
		\label{fig:Alibaba_per-minute_workload}
	\end{subfigure}
	\vspace{-0.1cm}
	\caption{(a) Original Alibaba per-day workloads. (b) Original Alibaba per-minute workloads. (c) Converted Alibaba per-day workloads. (d) Converted Alibaba per-minute workloads.}
	\label{fig:Alibaba_CPU}
\end{figure*}

To reduce the state space of our RL model, we process the workloads by dividing the workloads into a number of levels, e.g. using CPU utilization levels to represent the number of workloads, where the same scaling actions can be applied to the same level to reduce action space. For example, Fig. \ref{fig:Alibaba_CPU} shows the original continuous Alibaba’s workloads converted to 10 discrete CPU utilization levels at per-day and per-minute intervals, and each level represents 10\% utilization, e.g. level 0 represents utilization ranges from 0\% to 10\%.

\subsection{Neural Network-based Workload Predictor}
\label{sec:NNWP}

The Workload Predictor aims to accurately forecast the future workloads in system, and provides information for the RL-based Resource Scaler module to dynamically scale the number of pod replicates. The Workload Predictor module can be realized via different prediction approaches, such as ML-based prediction algorithms. \textit{ChainsFormer} considers the workloads prediction as a category of multi-variate time series forecast problem, where the workloads are time-relevant and multiple variables (e.g. CPU usage, memory usage, network throughput, and hard disk read/write) can influence the final prediction results. 

\textit{ChainsFormer} utilizes a GRU-based neural network validated in \cite{XuTOIT}, named esDNN, to predict future workloads, which can overcome the limitations of gradient explosion and disappearance when conducting long-term prediction. The esDNN can extract the key features of workloads, and convert multivariate time series forecasting into supervised learning to keep as much information as possible.
The performance of esDNN has been validated to achieve good accuracy in predicting workloads.

\subsection{Chains Analyzer}
\label{sec:chain_analyzer}
\begin{figure*}[t]
	\centering
	\begin{subfigure}[b]{0.45\linewidth}
		\centering
		\includegraphics[width=1\linewidth]{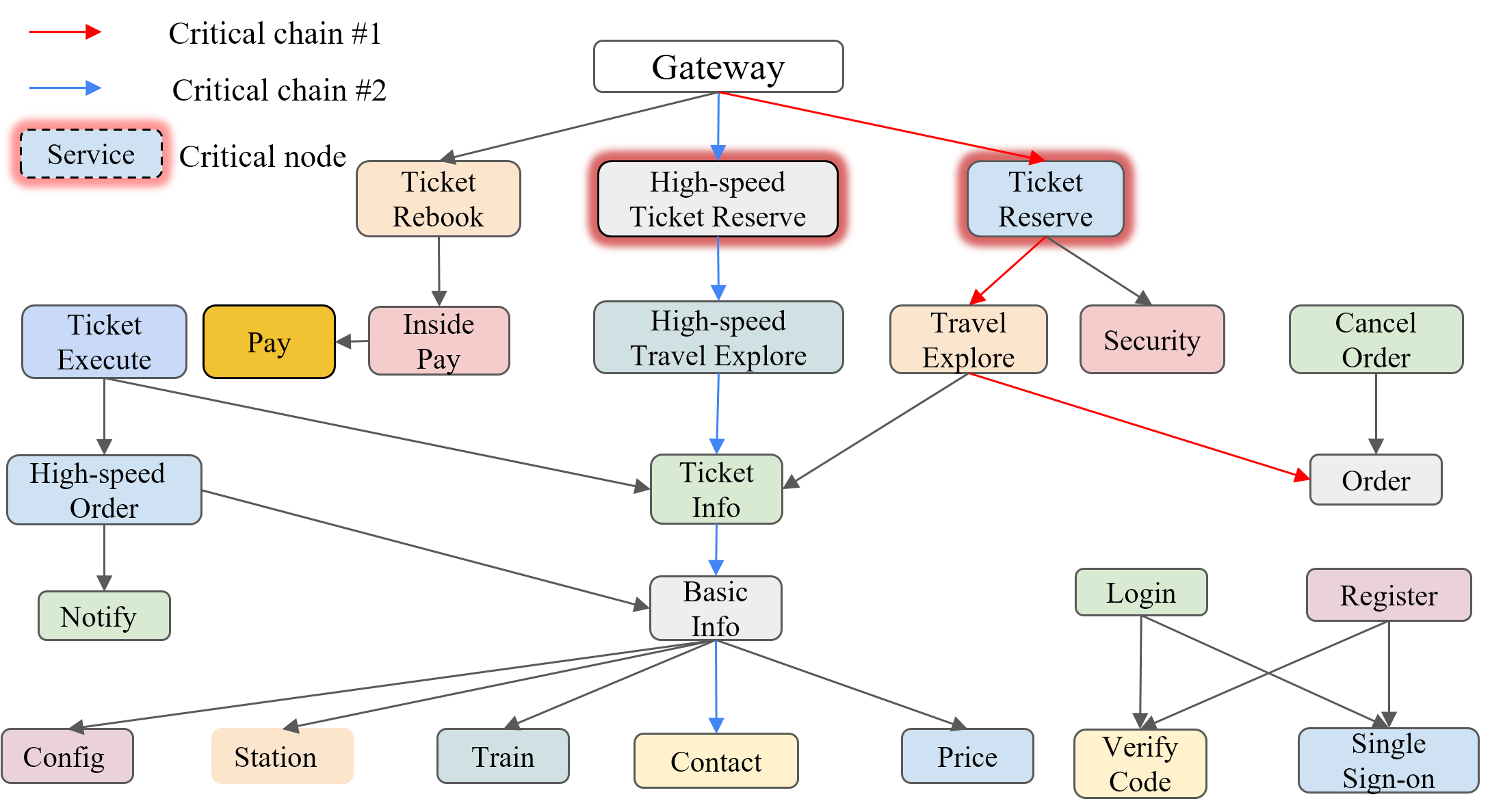}
		\caption{} 
		\label{fig:path}
	\end{subfigure}
	\begin{subfigure}[b]{0.45\linewidth}
		\centering
		\includegraphics[width=1\linewidth]{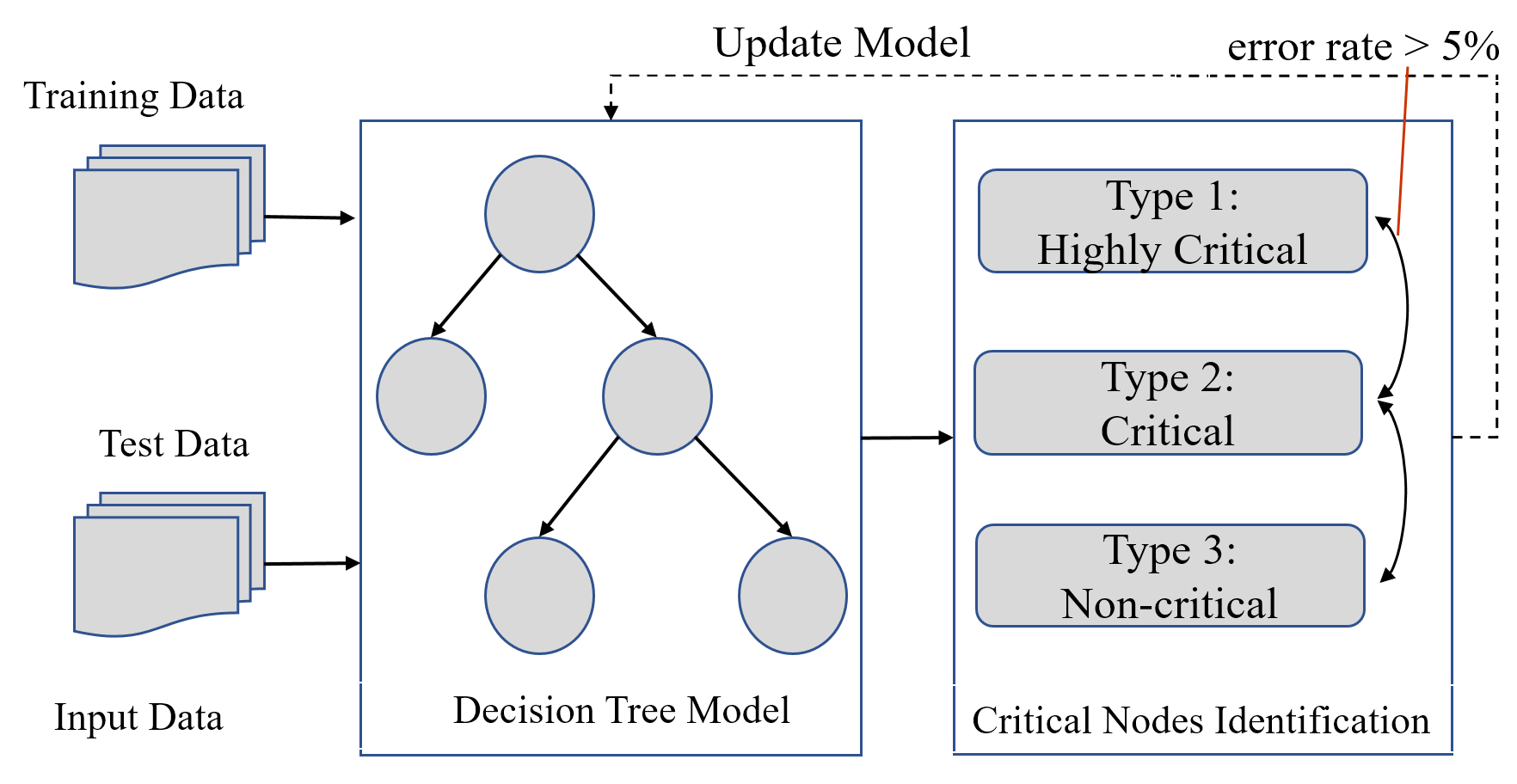}
		\caption{} 
		\label{fig:decision_tree}
	\end{subfigure}
	\vspace{-0.1cm}
	\caption{(a) Train-Ticket application with critical chain\protect\footnotemark. (b) Decision tree model for critical node identification}
	\label{fig:chain_analyzer}
\end{figure*}

One of the main goals of \textit{ChainsFormer} is to identify the critical chain efficiently and accurately based on tracing data and inter-dependencies, along with identifying the critical nodes that impact the latency of the critical chain. We define the critical chain as the one with the longest end-to-end latency, which represents the total time taken by a request to traverse the entire microservice chain, starting from the moment it enters the system until the user receives the response. Furthermore, the critical nodes (highlighted in Fig. \ref{fig:path} for Train-Ticket application) are defined as the nodes that have a substantial impact on the latency of the critical chain, and any performance degradation in these nodes can severely affect the performance of the microservices.

To identify the critical chain, \textit{ChainsFormer} uses tracing data to construct an execution graph that shows the processing sequence of a user request. The graph includes all the microservices involved in processing the request. \color{black}We then apply a weighted longest path algorithm \cite{ioannidou2013longest} to find the critical chain, which is the chain with the longest end-to-end latency. \color{black}The weight of each edge is the processing time between different nodes. This algorithm is lightweight and can adapt to changes in chains quickly. For example, if the blue chain in Fig. \ref{fig:path} has the highest latency, it can be identified as the critical chain. The critical chain will be changed to the red chain when its latency becomes to be the longest. We also identify critical nodes in the critical chain, which are the nodes that have a significant impact on latency. These critical nodes can significantly degrade the performance of microservices.

\color{black}The critical nodes are identified based on a decision tree, as shown in Fig. \ref{fig:decision_tree}. This tree classifies the nodes into critical and non-critical based on real-time data from the selected critical microservice chain and a trained model using historical running data. To reduce the overhead on the central node, the runtime data is stored on worker nodes and only fetched by the central node when model training or retraining is required. Nodes with high latency, CPU, and memory usage are more likely to be classified as critical nodes. In case the identification has a high error rate (e.g. 5\%), a model updating mechanism is triggered to update the decision tree. \color{black}

\footnotetext{https://github.com/FudanSELab/train-ticket}

\subsection{RL-based Resource Scaling}

The resource scaler uses RL techniques to determine the optimal scaling actions. Compared to static and meta-heuristic approaches, the RL-based approach can effectively explore a larger solution space and respond to dynamic status changes.
The RL-based resource scaler employs a hybrid scaling approach that includes both vertical scaling and horizontal scaling. Vertical scaling is used to quickly adjust resources such as CPU, memory, and network on the local machine, while horizontal scaling adds or removes active nodes in the system.

In \textit{ChainsFormer}, the problem of RL-based resource scaling is modeled as a Markov Decision Process \cite{WangTMC2021}. At each time interval $t$, the system state is represented by $s_t\in S$, and an action $a_t\in A$ can be taken to transition the state to $s_{t+1}$, yielding a reward of $R_{t+1}$ based on the policy $\pi_{\theta}$, which has configurable parameters $\theta$. The state space $S$ is associated with an action space $A$, and a transition matrix captures the probability of taking different actions during state transitions. The goal of RL is to optimize the policy to maximize the expected cumulative reward.

\color{black}To achieve this, \textit{ChainsFormer} employs the SARSA algorithm \cite{SaraTPDS2021} to learn the policy for the Markov decision process and estimate the expected cumulative reward of state-action pairs using the action-value function $Q_{t}(s,a)$. When action $a_t$ is taken at time interval $t$, the value of $Q_{t+1}$ is updated using the reward $R_{t+1}$ and propagated to the next time interval as: 
\begin{equation}
	\footnotesize
	Q_{t+1}(s_t, a_t) = Q_{t}(s_t, a_t) + \alpha[R_{t+1} + \gamma max_{a'}Q_{t}(s_{t+1}, a') - Q_{t}(s_t, a_t)],
\end{equation}
where $\alpha \in (0, 1]$ is the learning rate and $\gamma\in[0, 1]$ is the discount factor. 
To address the curse of dimensionality associated with updating the $Q$ Table with a large solution space, we train the model offline to minimize the loss function and reduce training time. Online training is used to make decisions and update actions with rewards. We also employed the divided load levels to reduce the state space, as discussed in Section \ref{sec:workloadsgen}. In addition, we use the SARSA algorithm to further reduce computational costs by using $R_{t+1} + \gamma max_{a'}Q_{t}(s_{t+1}, a')$ as the update target to guide the estimate of the true action-value function. This approach considers only the sampling of successive $s_{t+1}$, $a_{t+1}$, and immediate reward $R_{t+1}$. The estimation of the action-value function at the time interval $t+1$ is given by:
\begin{equation}
	\footnotesize
	Q_{t+1}(s_t, a_t) = Q_{t}(s_t, a_t) + \alpha[R_{t+1} + \gamma Q_{t}(s_{t+1}, a_{t+1}) - Q_{t}(s_t, a_t)],
\end{equation}
where the $Q_t(s_{t+1}, a_{a+1})$ and each update can be obtained via one-step transition $(s_t, a_t, R_{t+1}, s_{t+1}, a_{t+1})$ of the state–action–reward–state–action pair. \color{black}


To implement the RL-based resource scaler module, we utilize various parameters of the current pod as inputs to the RL model. These parameters include the load state, the position of the pod in the chain, and the latency of the microservice. The RL model considers the state $s_t\in S$ to represent the current status of the microservice chain, and action $a_t\in A$ comprises scaling operations that adjust the chain status and provisioned resources by a specific amount.
We also assume the presence of a set of physical machines $P=(M_1,M_2,\ldots,M_K)$ in the system that provides resources. Each physical machine $M_k$ is represented by a tuple $U_k=(u_k^1, u_k^2, \ldots, u_k^I)$, where $u_k^i$ represents the resource utilization of type $i$ out of a total of $I$ resource types on physical machine $M_k$. For each $M_k$, we denote the set of possible actions as $a_k^i=\{h_k, v_k^i\}$, where $h_k\in[-n,n]$ represents the number of horizontal replicates that can be added or removed, $v_k^i\in[-m,m]$ represents the amount of vertical scaling that can be applied to resource type $i$. A positive value of $h_k$ or $v_k^i$ indicates that more resources are added, whereas negative values indicate resource removal. Given $K$ as the total number of physical machines, the final set of actions is represented as the Cartesian product of the sub-action sets: $A= \prod_{k=1}^K \prod_{i=1}^I a_k^i$.

The main objective of the \textit{ChainsFormer} system is to enhance resource utilization while ensuring QoS requirements are met. Therefore, the reward function is designed to consider two key metrics: resource utilization and response time. The reward for resource utilization is formulated in Eq.~\eqref{eq:rewardutil}.
\begin{equation}
	\label{eq:rewardutil}
	R_{u}(u_k) = \begin{cases}
		\frac{\sum_{k=1}^K U_k^{max}-u_k}{K} + 1, & u_k \leq U_k^{max}, \\
		\frac{\sum_{k=1}^K u_k - U_k^{max}}{K} + 1, & u_k > U_k^{max},
	\end{cases}
\end{equation}
where $U_k^{max}$ represents the highest utilization threshold of all resource types for physical machine $M_k$, and $u_k$ is the current utilization of $M_k$. The system receives a positive reward when the utilization is below the threshold, and the reward decreases when the utilization is higher or significantly lower than the predefined threshold.

\color{black}The reward for response time, denoted as $R_{q}(rt)$, is modeled based on the maximum acceptable response time $RT_{max}$. 
\begin{equation}
	\label{eq:reward}
	R_{q}(rt) = \begin{cases}
		e^-({\frac{rt-RT_{max}}{RT_{max}})^2}, & rt >  RT_{max}, \\
		1, & rt \leq RT_{max},
	\end{cases}
\end{equation}
which shows that when the system is operating normally, the reward is 1. However, as the system's performance degrades and violates the $RT_{max}$, the reward gradually decreases and converges to 0. \color{black}		

\color{black}The final reward value is based on the resource utilization $R_{u}$ and response time $R_{q}$ at time interval $t$, which is formulated as follows:
\begin{equation}
	\label{eq:finalreward}
	r(s_t, a_t) = \frac{R_{q}^t}{R_{u}^t},
\end{equation}
where higher values of $R_{q}^t$ and lower values of $R_{u}^t$ can increase the total reward.\color{black}	

\begin{algorithm}[t]
	\label{alg:general}
	\color{black}
	\footnotesize
	\caption{\textit{ChainsFormer}: Overall Procedure} 
	\label{alg:ChainsFormer_general}
	\SetAlgoLined
	\SetKwInOut{Input}{Input}\SetKwInOut{Output}{Output}
	\SetKwFor{ForAll}{forall }{do}{end forall}
	\Input{Table $Q(s, a)$ contains all state/action pairs from  experience pool by offline training, time intervals $T$, probability of random action $\epsilon$, learning rate $\alpha$, discount factor $\gamma$}
	
	Initialize system status, and monitoring model;
	
	\For{$t$ from $1$ to $T$}{
		
		$U_t^k$  $\leftarrow$ Resource utilization of $M_k$ at time interval $t$;
		
		$W_{t-1} $ $\leftarrow$ Workloads level at time interval $t-1$;
		
		$\hat{W_t}$  $\leftarrow$  Predicted workload level;
		
		\If{$\hat{W_t} \neq W_{t-1}  $ }{
			Choose a action from action set $A$ with $\epsilon$ probability, or select an action with the $\max(Q_t(s_t, a_t))$;
			
			Conduct $a_t$ =\{$h_k(t), v_k^i(t)$\} with horizontal scaling and vertical scaling
			
			\If{online training is triggered}{
				
				$s_{t+1}$ $\leftarrow$ system state at time interval $t+1$;
				
				$R_{t+1}$ $\leftarrow$  reward calculation by Equation (\ref{eq:finalreward});
				
				Update $Q$ value:  $Q_{t+1}(s_t, a_t) = Q_{t}(s_t, a_t) + \alpha[R_{t+1} + \gamma Q_{t}(s_{t+1}, a_{t+1}) - Q_{t}(s_t, a_t)];$
				
			}
			Store transition $(s_t, a_t, R_{t+1}, s_{t+1}, a_{t+1})$in experience pool;
			
		}
		
	}
	
\end{algorithm}

Algorithm \ref{alg:ChainsFormer_general} outlines the overall procedure of \textit{ChainsFormer}. Initially, the algorithm collects the system status to enable the RL process (line 1), which includes monitoring the workloads level, resource utilization, and metrics at each time interval to construct the complete system states (lines 3-5). Upon a change in workload level (line 6), resources are dynamically scaled to optimize resource usage while maintaining the required QoS. The SARSA algorithm  commences by selecting actions randomly with a probability of $\epsilon$ from the experience pool and transitions to another state (line 7). The chosen actions entail vertical and horizontal scaling to allocate resources effectively (line 8).
\textit{ChainsFormer} facilitates online training by storing the transition $(s_t, a_t,R_{t+1}, s_{t+1}, a_{t+1})$ in the experience pool and subsequently updating the decisions based on rewards with better outcomes (lines 9-14).

\section{Performance Evaluations}
\label{sec:performance}

\subsection{Experimental Settings}

We use the workload dataset provided by Alibaba\footnote{Alibaba Cluster Trace Program: https://github.com/alibaba/clusterdata/tree/v2018} as demonstrated in Section \ref{sec:workloadsgen}, which includes 8-day data traces from homogeneous 4,034 servers. We utilize the Locust toolkit to generate resource usage based on profiled data of machines. We evaluate the performance of the Train-Ticket application (a larger application than the Sock Shop used for motivation in Section \ref{sec:Policy}) and use the Jaeger monitoring toolkit to track the distribution of requests. The application is deployed on a Kubernetes-based cluster consisting of five nodes, each with an Intel Xeon E5-2660 processor and 64 GB of RAM. One physical machine serves as the master of the cluster, while the others serve as workers.

\subsection{Baselines and Metrics}

We have compared \textit{ChainsFormer} (\textbf{CF}) with 3 state-of-the-art baselines implemented by us. 

\textbf{KS \cite{burns2019kubernetes}}: it is employed by native Kubernetes and mainly relies on horizontal scaling, which involves dynamically adding or removing the number of replicas. It follows a threshold-based approach based on resource usage metrics such as CPU and memory, where more replicas are added when the pre-defined resource threshold is exceeded (e.g. CPU utilization > 0.7) and vice versa.




\textbf{AUTO \cite{rzadca2020autopilot}}: it is derived from Google Autopilot and uses a hybrid approach to scale resources based on workloads. The approach combines horizontal scaling and vertical scaling to dynamically adjust the allocated resources to tasks based on historical data. 

\textbf{FIRM \cite{qiu2020firm}}:  it utilizes machine learning techniques, specifically support vector machine and reinforcement learning, to identify and mitigate microservices responsible for SLO violations.

\color{black}We have adopted three widely used metrics to evaluate the performance:
\textit{1) Requests per second (RPS)}  represents the system's ability to process requests within a specific time period, and a higher value shows better performance.
\textit{2) Number of failures} indicates the number of requests that were not processed or did not receive a response due to an overloaded situation. A lower value for this metric represents a more reliable system. 
\textit{3) Average response time: } is a dominant metric to measure performance, and a good autoscaling algorithm should aim to reduce it.\color{black}

\subsection{Experiment Analyses}
\label{sec:discussion}

\begin{figure}[t]
	\centering
	\begin{subfigure}{0.3\linewidth}
		\centering
		\includegraphics[width=0.99\linewidth]{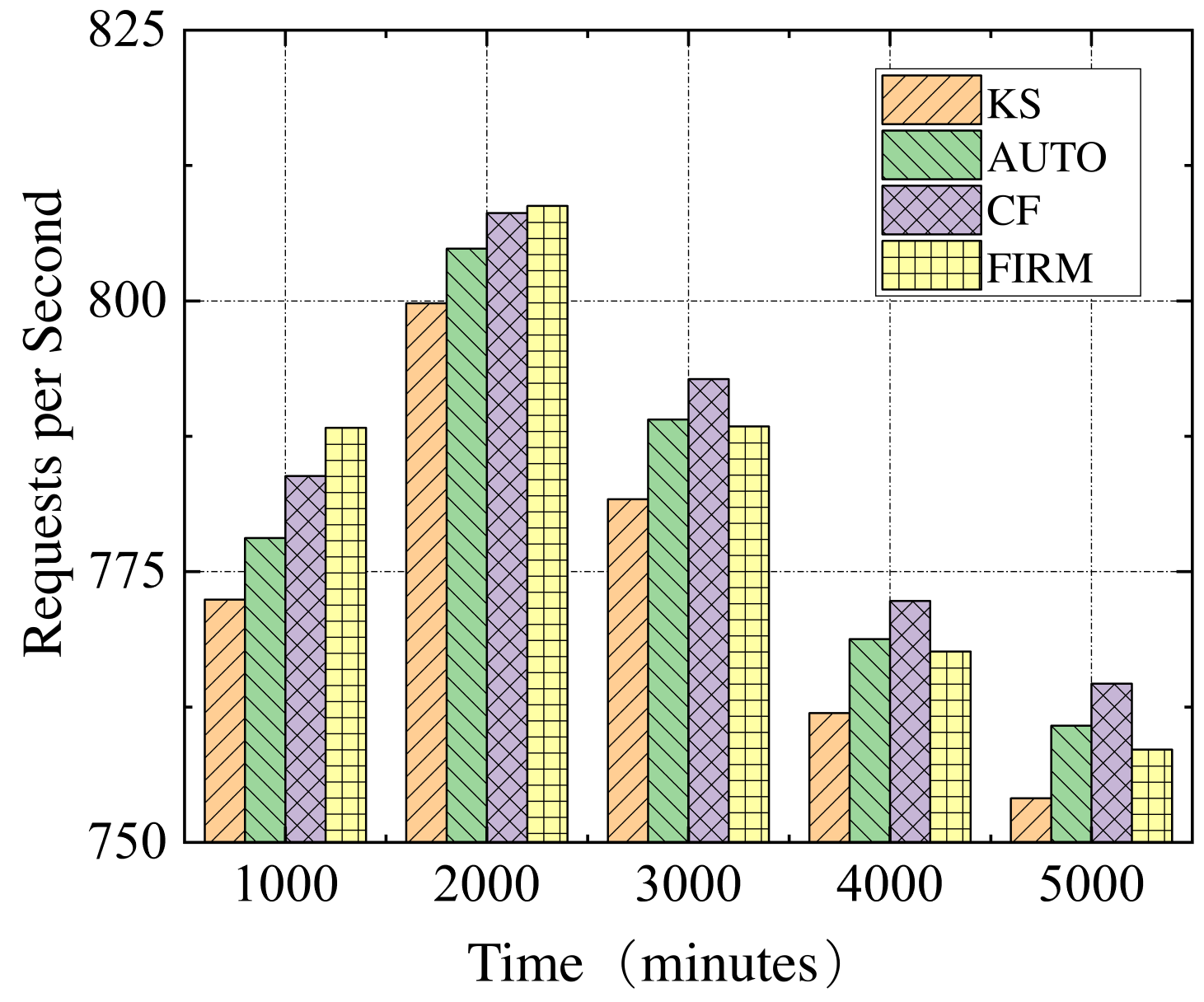}
		\caption{}
		\label{fig:RPS}
	\end{subfigure}
	\begin{subfigure}{0.3\linewidth}
		\centering
		\includegraphics[width=0.99\linewidth]{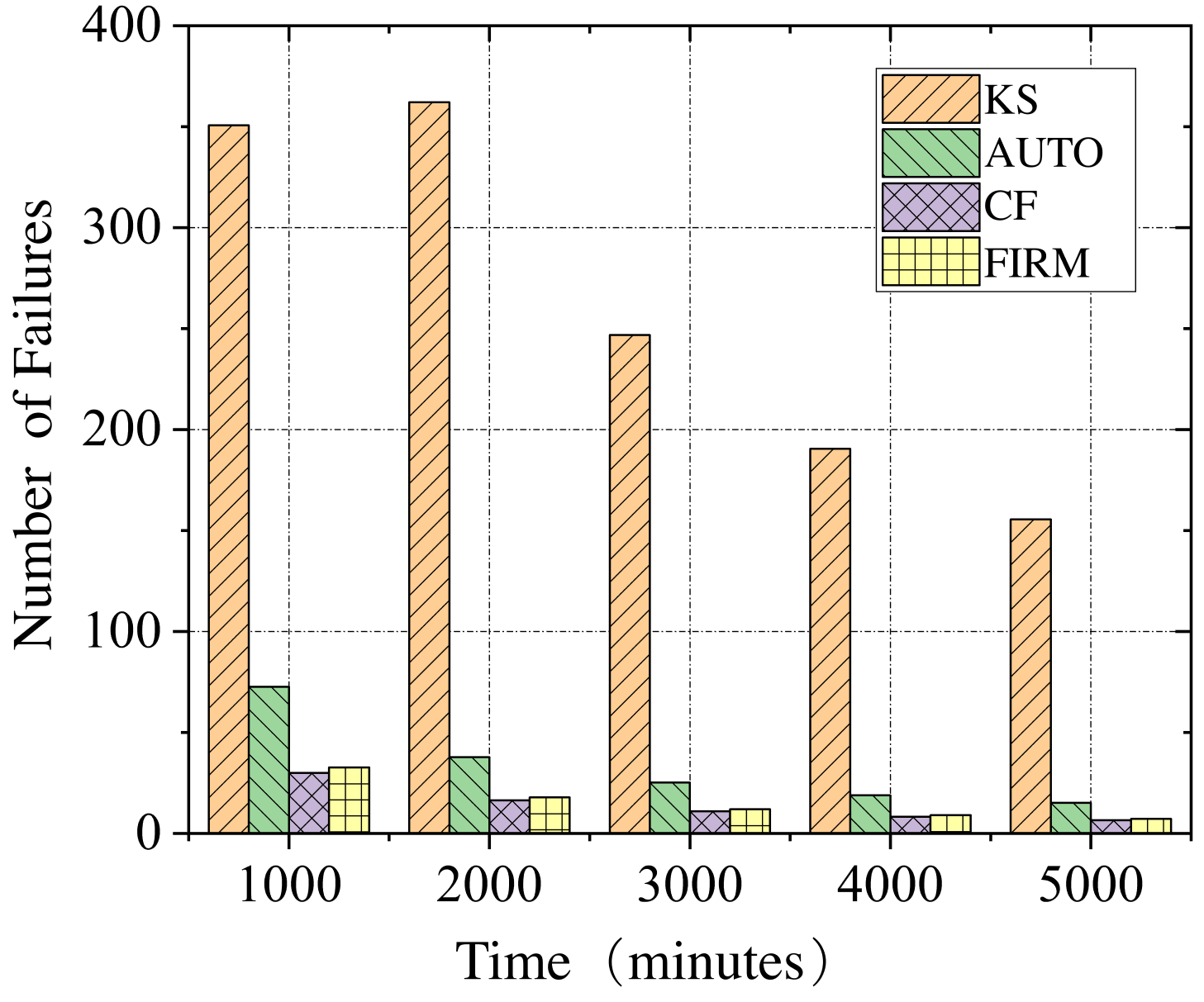}
		\caption{}
		\label{fig:FC}
	\end{subfigure}
	\begin{subfigure}{0.3\linewidth}
		\includegraphics[width=0.99\linewidth]{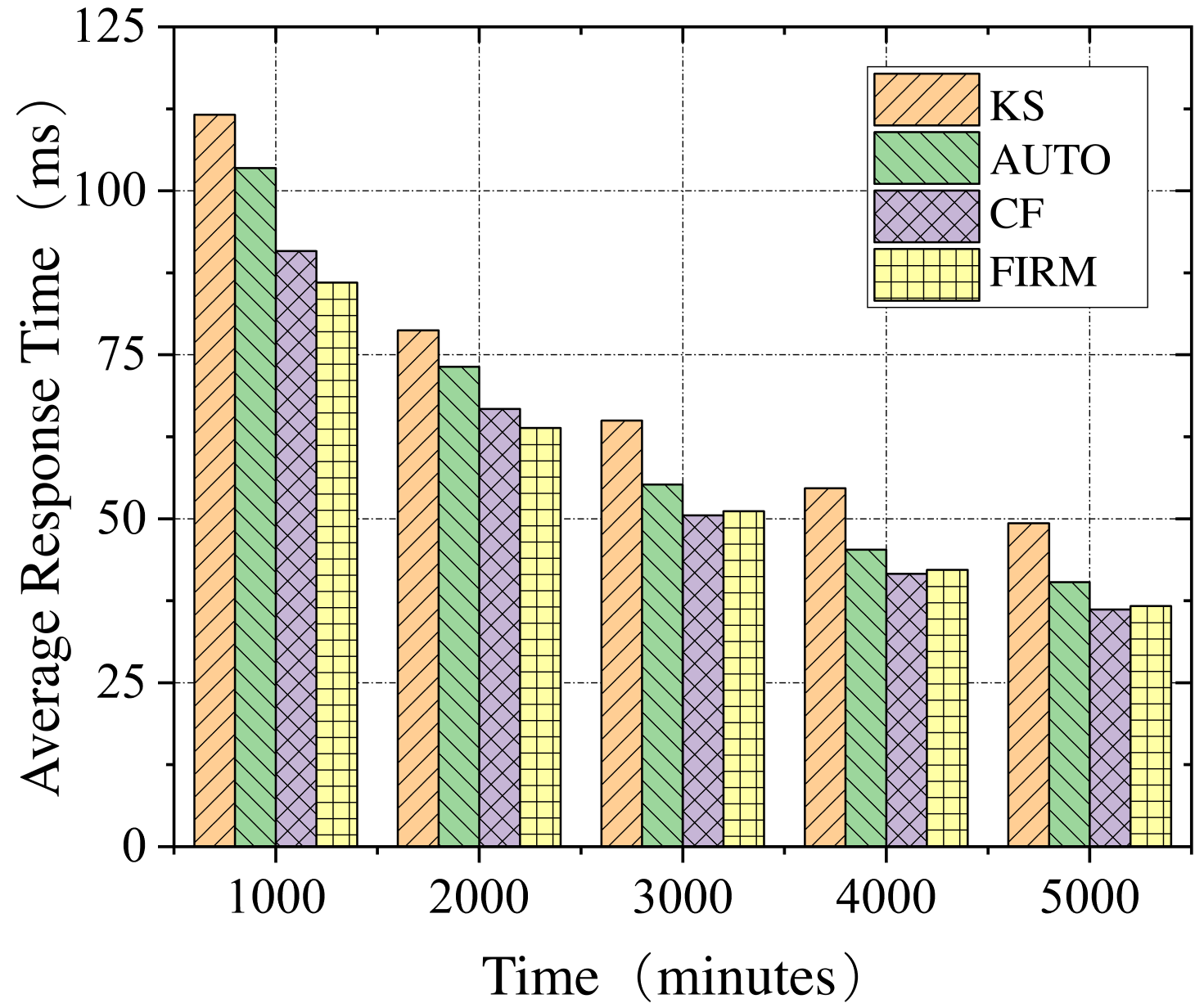}
		\caption{}
		\label{fig:ART}
	\end{subfigure}
	\vspace{-0.2cm}
	\caption{Comparison of (a) requests per second, (b) number of failures, and (c) average response time. }\label{fig:act}
\end{figure}

Due to page limitations, we present key results. Fig. \ref{fig:RPS} compares the average requests per second over different time periods. To highlight differences among periods, we analyze results over 5 periods (e.g. 1,000 minutes, 2,000 minutes, and 5,000 minutes), covering short-term and long-term comparisons. It is noteworthy that the loads significantly vary during different time periods. For instance, the highest loads were observed during the first 2,000 minutes, and the average load during the 5,000 minutes period was much lower.
It is observed that the KS approach performs the worst in terms of RPS compared to other baselines. This could be due to the limited capability of the threshold-based approach. The AUTO approach, which leverages ML-based techniques, can process larger RPS compared to KS. The FIRM approach can obtain better RPS during the first 3,000 minutes, but during the 4,000-5,000 minutes, it performs worse than AUTO. Our proposed approach, \textit{ChainsFormer}, can achieve the best RPS in the long-term, i.e., when the time period is larger than 3,000 minutes. This optimization comes from our more accurate identification of critical chains and nodes.
At the early stage of request processing, FIRM performs well when the critical path is identified. However, after load changes, the identified critical path may not be critical anymore. Additionally, it is reasonable to note that FIRM with a static critical path does not fit workloads with high variances well. In conclusion, CF optimized the requests per second up to 8.1\% compared to the baselines.

Fig. \ref{fig:FC} illustrates the comparison of the number of failures, presenting the average results in five different time periods. It is observed that KS has the highest number of failures compared to other baselines due to its static policy, which shows that it struggles to handle high-variant workloads. AUTO significantly reduces the number of failures. For instance, during the first 1,000 minutes, AUTO reduces the failures from 350 to 80 by leveraging historical data. FIRM and CF further optimize the failures by utilizing critical chains and nodes, where the results are quite close. Overall, CF can reduce the number of failures by 8.3\% compared to FIRM.


Fig. \ref{fig:ART} depicts the comparison of average response time. Among all five time periods, the average response time of KS is at the highest value, which we consider as a benchmark test to analyze the performance of the other three algorithms. AUTO is optimized compared to KS and maintains the second-highest response time. It optimizes around 10\% of response time, for example, decreasing from 110 ms to 100 ms during the first 1,000 minutes. The results of CF and FIRM are consistent with the analyses of RPS. In the early stage, FIRM slightly outperforms CF, while in the long run (e.g. 3,000 to 5,000 minutes), CF achieves a lower response time than FIRM. Overall, \textit{ChainsFormer} optimizes response time by 1.4\% to 26.6\% compared to the baselines.

\begin{figure}[t]
	\centering
	\begin{subfigure}{0.4\linewidth}
		\centering
		\includegraphics[width=0.8\linewidth]{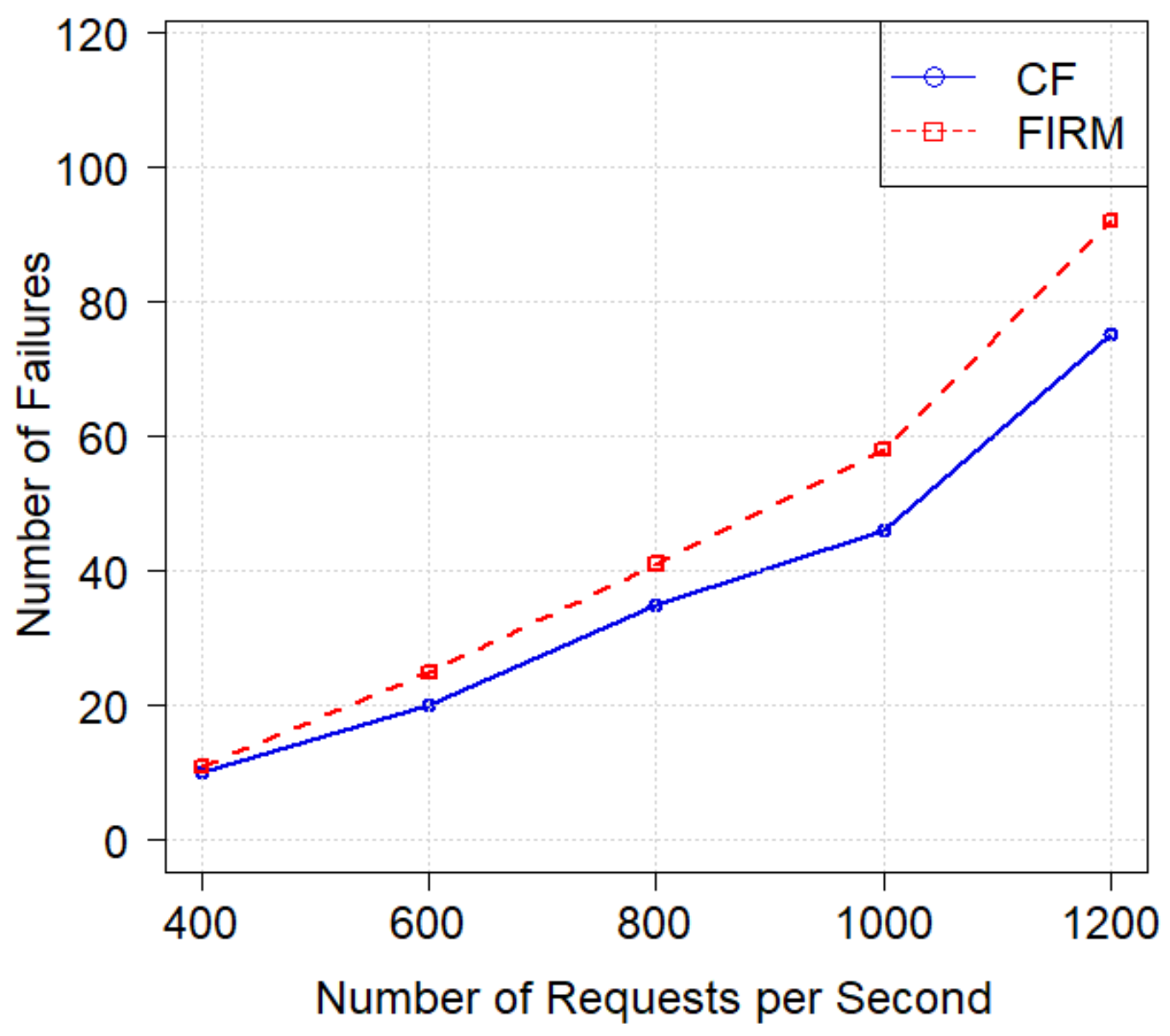}
		\caption{}
		\label{fig:scalability_faliures}
	\end{subfigure}
	\begin{subfigure}{0.4\linewidth}
		\centering
		\includegraphics[width=0.8\linewidth]{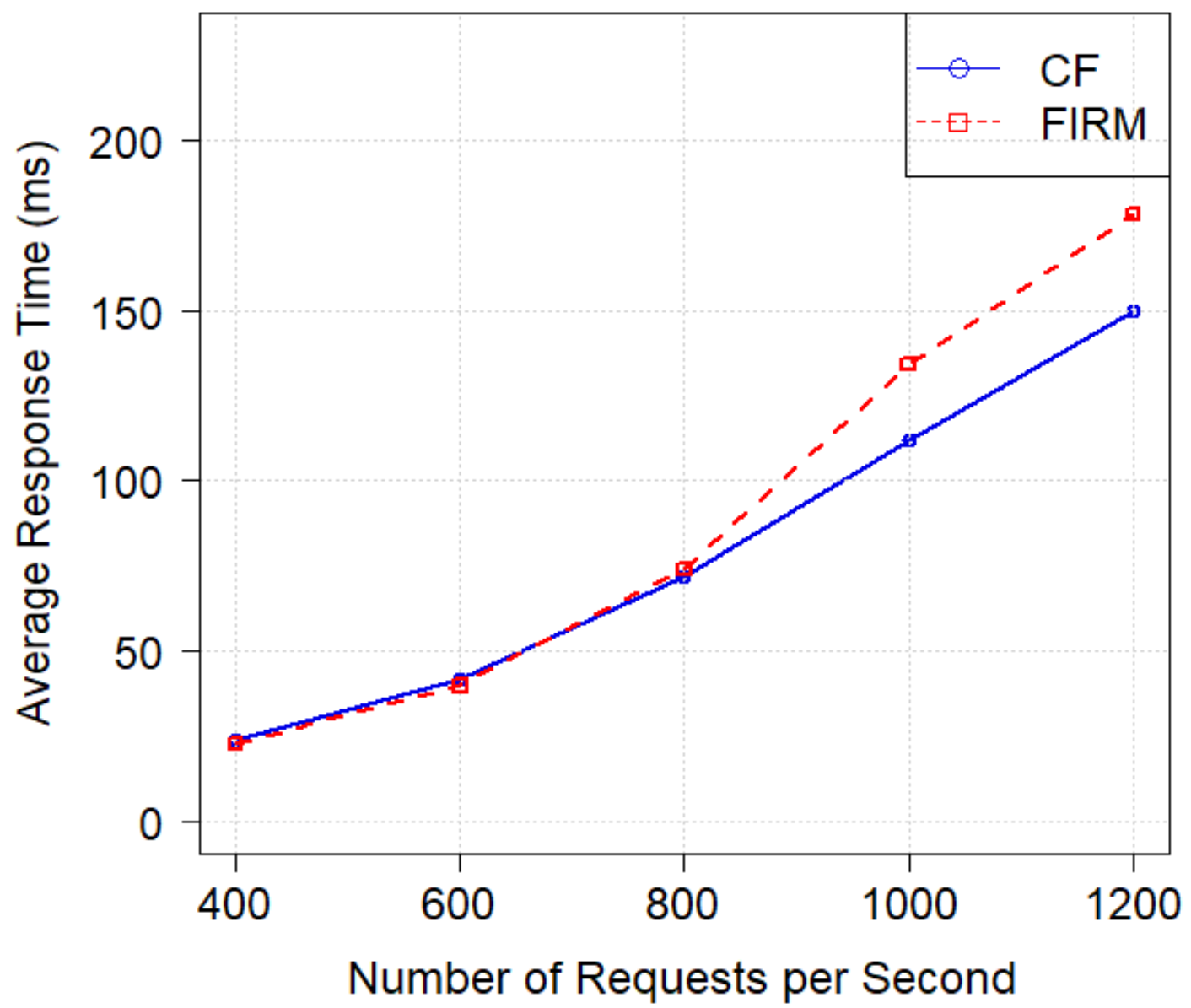}
		\caption{}
		\label{fig:scalability_requests}
	\end{subfigure}
	\vspace{-0.2cm}
	\caption{\color{black}Scalability comparison of (a) number of failures, and (b) average response time when the number of requests increase.} \label{fig:scalability}
\end{figure}

\color{black}To evaluate the scalability of \textit{ChainsFormer}, we conducted experiments comparing it with FIRM under different numbers of requests as shown in Fig. \ref{fig:scalability}.  
We gradually increased the number of requests (from 400 to 1200 per second) and monitored the system's performance. The number of failures in \textit{ChainsFormer} exhibited a slower rate of increase compared to FIRM as the number of requests grew. 
Similarly, the average response time in \textit{ChainsFormer} remained relatively stable as the number of requests grew. In contrast, FIRM experienced a more pronounced increase in average response time under the same conditions. 
These findings validate the scalability of \textit{ChainsFormer} and its ability to handle larger workloads while maintaining good performance. The results suggest that \textit{ChainsFormer} is a promising solution for scaling microservice-based systems in scenarios with dynamic and growing request loads.\color{black}

\section{Conclusions}
\label{sec:conclusion}

In this paper, we propose \textit{ChainsFormer}, a microservice scaling approach that combines deep learning and reinforcement learning techniques to dynamically adjust resource allocation based on workload predictions and critical chain identification. By leveraging decision trees for rapid identification of critical chains and nodes, and using reinforcement learning to make real-time scaling decisions, \textit{ChainsFormer} optimizes resource usage while maintaining high-quality of service in terms of response time, number of failures, and requests per second. Our experiments, conducted on a representative microservices application, show that \textit{ChainsFormer} outperforms state-of-the-art algorithms from research and industry in terms of QoS optimization. Our approach has the potential to significantly improve the efficiency and reliability of microservices-based applications in cloud computing environments.

\noindent{
\textbf{Acknowledgments}.\addcontentsline{toc}{section}{Acknowledgment} 
This work is supported by National Key R \& D Program of China (No.2021YFB3300200), the National Natural Science Foundation of China (No. 62072451, 62102408), Shenzhen Industrial Application Projects of undertaking the National key R \& D Program of China (No. CJGJZD20210408091600002), Shenzhen Science and Technology Program (No. RCBS20210609104609044), and Alibaba
Group through Alibaba Innovative Research Program.  }

		\bibliographystyle{splncs04}
		\bibliography{references}

	\end{sloppypar}
\end{document}